# Video Encryption: A Survey

Jolly shah and Dr. Vikas Saxena

Dept. of CSE & IT, Jaypee Institute of Information Technology
Noida, Uttar Pradesh 201307, India

Dept. of CSE & IT, Jaypee Institute of Information Technology
Noida, Uttar Pradesh 201307, India

**Abstract**
Multimedia data security is becoming important with the continuous increase of digital communications on internet. The encryption algorithms developed to secure text data are not suitable for multimedia application because of the large data size and real time constraint. In this paper, classification and description of various video encryption algorithms are presented. Analysis and Comparison of these algorithms with respect to various parameters like visual degradation, encryption ratio, speed, compression friendliness, format compliance and cryptographic security is presented.

## 1. Introduction

The wide use of digital images and videos in various applications brings serious attention to security and privacy issues today. Data encryption is a suitable method to protect data. Till now, various encryption algorithms have been proposed and widely used (DES, RSA, IDEA, AES etc.), most of which are used for text and binary data. It is difficult to use them directly in video encryption as video data are often of large volumes and require real time operations. In past decade, some video encryption algorithms have been reported, most of which are based on MPEG ½ codec [2] [3].

In practical applications, for a video encryption algorithm, security, time efficiency, format compliance and compression friendliness are really important [1]. Among them, security is the basic requirement, which means that the cost of breaking the encryption algorithm is no smaller than the ones buying the video's authorization. The time efficiency means encryption and decryption should not take much time as heavy delay is not acceptable in real time. Apart from that encryption and decryption should not affect compression ratio. The format compliance means that the encryption process does not change the encoded bit stream's data format in order to support such direct operations as browsing, playing, cutting, copying and so on.

In this paper, classification of different video encryption algorithms is presented. The paper is structured as follows: In section 2, classification and performance parameters of video encryption algorithms are given. In section 3, description of these algorithms as per classification is presented. Section 4 provides comparison of these algorithms and finally conclusion is drawn in section 5.

## 2. Classification and Performance Parameters

In section 2.1, classification of video encryption algorithms is presented. Video encryption algorithms are classified based on their unique way of encrypting data. Section 2.2 presents performance parameters of these algorithms based on which evaluation and comparison is done.

2.1 Classification of Video Encryption Algorithms

We classify video encryption algorithms in four categories.

**Fully layered Encryption**: In this class, whole content of video is first compressed and then encrypted using standard tradition algorithms like DES, RSA, IDEA, AES etc. This technique is not suitable in real time video applications due to heavy computation and slow speed.

**Permutation based Encryption**: The video encryption algorithms in this class mainly use different permutation algorithms to scramble or encrypt the video contents. It is not necessary to scramble each and every byte. Some algorithms use permutation list as secret key to encrypt video contents.

**Selective Encryption**: The algorithms in this class selectively encrypt the bytes within video frames. As these algorithms are not encrypting each and every byte of video data, it reduces computational complexity.

**Perceptual Encryption:** Perceptual encryption requires that quality of aural/visual data is only partially degraded by encryption i.e., the encrypted multimedia data are still partially perceptible after encryption. In this aural/visual quality degradation can be continuously controlled by a factor p.

## 2.2 Performance Parameters

We need to define a set of parameters based on which we can evaluate and compare video encryption algorithms. Some parameters listed below are gathered from literature.

**Visual Degradation (VD):** This criterion measures the perceptual distortion of the video data with respect to the plain video. In some applications, it could be desirable to achieve enough visual degradation, so that an attacker would still understand the content but prefer to pay to access the unencrypted content. However, for sensitive data, high visual degradation could be desirable to completely disguise the visual content.

**Encryption ratio (ER):** This criterion measures the ratio between the size of encrypted part and the whole data size. Encryption ratio has to be minimized to reduce computational complexity.

**Speed (S):** In many real-time video applications, it is important that the encryption and decryption algorithms are fast enough to meet real time requirements.

**Compression Friendliness (CF):** An encryption algorithm is considered compression friendly if it has no or very little impact on data compression efficiency. Some encryption algorithms impact data compressibility or introduce additional data that is necessary for decryption. It is desirable that size of encrypted data shoud not increase.

**Format Compliance (FC):** The encrypted bit stream should be compliant with the compressor. And standard decoder should be able to decode the encrypted bit stream without decryption.

**Cryptographic Security (CS):** Cryptographic security defines whether encryption algorithm is secure against brute force and different plaintext-ciphertext attack? For highly valuable multimedia application, it is really important that the encryption algorithm should satisfied cryptographic security.

## 3. Video Encryption Algorithms

Videos are transferred through various types of computer network. To secure video communication different encryption methodologies are used. Due to huge size of digital videos they are generally transmitted in compressed formats such as MPEG 1/2/4[4][5][7], H.263/ H.264/AVC [6][7]. Various encryption algorithms have been proposed in literature. This section presents detail description of these algorithms.

### 3.1 Fully layered Encryption

In this scheme, the whole content is first compressed. Then, the compressed bitstream is entirely encrypted using a standard cipher DES [8] or AES [9].

#### 3.1.1 Naïve Technique

The most straightforward method to encrypt every byte in the whole Moving Picture Expert Group (MPEG) stream using standard encryption schemes such as DES or AES. The idea of naïve algorithm is to treat the MPEG bitstream as text data and does not use any of the special structure [2]. It provides the security to whole MPEG stream because every byte is encrypted, and no algorithm exists to break triple DES or AES so far. It is not applicable solution for big video, because it is very slow especially when we use triple DES. Because of the encryption operation delay increases and overload will be unacceptable for real time video application. As encryption is performed after compression, no impact is observed on compression efficiency.

### 3.2 Permutation based Encryption

This section discusses about the encryption algorithms which works mainly on achieving visual degradation on permutation principle.

#### 3.2.1 Pure Permutation

The idea of pure permutation algorithm is simply scrambles bytes within a frame of MPEG stream by permutation. It is extremely useful in situation where the hardware decodes the video, but decryption must be done in software. Adam J. slagell demonstrates that pure permutation algorithm is vulnerable to known-plaintext attack, and hence its use should be careful considered. [10], because by comparing the ciphertext with the known frames, the adversary could easily figure out the secret permutation list. Once the permutation list is figured out, all frames could be easily decrypted.

#### 3.2.2 Zig-Zag Permutation

In Zig-Zag permutation [11],instead of mapping the 8X8 block to 1X64 vector in "Zig-Zag" order, it maps individual 8x8 block to 1x64 vector by using a random permutation list (secret key). There are many ways to produce a permutation list which has uniform distribution over all possible permutations. This algorithm consists of three steps.

i) Generate a permutation list with cardinality 64.
ii) Complete splitting procedure after 8x8 block is quantized.
iii) Apply the random permutation list to the split block, and pass the result to the entropy coding procedure.

Since mapping Zig-Zag order and mapping according to random permutation list have the same computational complexity, the encryption and decryption add very little overhead to the video compression and decompression processes. However, this method decreases the video compression rate because the random permutation distorts the probability distribution of Discrete Cosine Transform (DCT) coefficients and make Huffman table used less than optimal. Zig- Zag permutation algorithm cannot withstand the known-plaintext attack. By assuming that we know certain frames of video in advance the secret key can be easily figure out by simply comparing known plaintext attack with corresponding encrypted frame. To solve this problem, binary coin flipping sequence method together with two permutation lists is used. For each 8x8 block a coin is flipped. If it is a tail, the permutation list 1 (key1) is applied to block. If it is a head, the permutation list 2 (key2) is applied to the block. This method is vulnerable to the ciphertext only attack, because non zero AC coefficients have the tendency to gather in the upper left corner of the block, it would be easy for an adversary to determine which key is used.

### 3.2.3 Huffman Codeword Permutation

It is a lightweight mpeg video encryption which incorporates encryption with MPEG compression in one step [12]. The primary goal of this methodology is to save computation time by taking the advantage of combining MPEG compression and data encryption and at the same time avoid decreasing video compression rate. In this permutation, Huffman codeword list is used as a secret key. During MPEG encoding, the encoder uses the secret key instead of standard Huffman codeword list. Since MPEG compression rate depends on Huffman codeword list, if we use an arbitrarily Huffman codeword list to encode the MPEG video, the compression rate may decrease. To avoid affecting compression rate, it limits the permutation of Huffman codeword list (secret key) to those codewords which have the same length as the standard Huffman codeword. Second, it seems that not all of permutations of the Huffman codeword list can be used as an encryption keys. This makes key generation difficult since a generated key has to be tested for validity before using.

### 3.2.4 Compression Logic based Random Permutation

The proposed algorithm is Compression logic based video encryption algorithm [13]. Instead of randomly permuting 8x8 coefficients of a single DCT block, the random permutation is applied to a number of permutation groups. Each permutation group contains the DCT coefficients of same frequency from every single block of a frame, regardless of I,P or B frame. Obviously, since each DCT block has 64 coefficients frequencies so that 64 permutation groups can be formed, the proposed algorithm runs random permutations on each of the permutation groups to encrypt a single video frame. After the random permutation the encrypted video data is compressed by standard RLE. It is also a selective algorithm since only a small number of permutation groups can be encrypted based on the requirements of confidentiality. It is reliable against brute force attacks due to a very large key space. It is secure against DCT vulnerability.

### 3.2.5 Correlation Preserving Permutation

Most encryption algorithms have a randomization effect on the source data, and cannot be effectively applied before compression stage. Using correlation preserving permutation [14] one can perform encryption prior to video encoding. In this scheme, Sorted, as well as "almost sorted" frames are strongly spatial correlated. Such permuted frames are in many instances even more compressible in terms of spatial only coding than the original source frames. When a sorting permutation of previous frame acts on the current frame, it produces what we refer to as an "almost sorted" frame. Transmitting a compressed frame from which the initial permutation can be computed is efficient. Once an initial permutation is transmitted through a secure channel, the sender uses it to "almost sort" the next frame. It is shown that except in rare circumstances, a sorted or "almost sorted" frame can be safely sent through the regular, non secure channel. By calculating a sorting permutation of the received frame, the receiver uses it to recover the next frame. There is no secret key on which a permutation is generated. This method relies on the sorting permutation of previous frame, and thus, a key is directly dependent on the plaintext. Under a chosen plaintext attack, the adversary can compute the sorting permutation for the chosen frame, but this gives no information about the sorting permutations for the unknown frames. The limited known-plaintext attack is applicable to our method, because the adversary can recover all frames that follow the known frame until the scene changes and key frame is updated.

### 3.3 Selective Encryption

In traditional video protection schemes, called fully layered, the whole content is first compressed and then compressed bit stream is entirely encrypted using a standard cipher. This scheme is unsuitable in real time application due to high delay and computation complexity. This section discusses about selective encryption which only encrypt a subset of the data. The aim of selective encryption is to reduce the amount of data to encrypt while preserving a sufficient level of security.

### 3.3.1 Methodology proposed by Meyer and Gadegast

This methodology is proposed for MPEG videos [15]. This method uses traditional encryption methods RSA or DES in CBC mode to encrypt MPEG video stream. It implements 4 level of security. (i) Encrypting all stream headers. (ii) Encrypting all stream headers and all DC and lower AC coefficients of intracoded blocks. (iii) Encrypting I-frames and all I-blocks in P- and B frames. (iv) Encrypting all the bit streams. The number of I blocks in P or B frames can be of the same order as the number of I blocks in I frames. This reduces considerably the efficiency of the selective encryption scheme [16]. Encryption ratio may vary based on which parameters are encrypted. Encrypting only headers have very less encryption ratio. But encrypting all the bitstreams have 100% encryption ratio. Speed of this methodology again varies based on traditional algorithm in use such as DES or RSA and number of parameters that are encrypted. Many security levels can be obtained. Encrypting only stream headers is not sufficient since this part is easily predictable. But encrypting all the bit streams can provide high security. Detailed cryptanalysis of this methodology is not defined. A special encoder and decoder are required to read unencrypted SECMPEG stream. The encoder proposed is not MPEG compliant.

### 3.3.2 Methodology proposed by Spanos and Maples

Aegis mechanism is proposed in [17]; it encrypts intraframes, video stream header and the ISO 32 bits end code of the MPEG stream using DES in CBC mode. Experimental results were conducted by the authors showing the importance of selective encryption in high bitrate video transmission to achieve acceptable end-to-end delay. It is also shown that full encryption creates bottleneck in high bitrate distributed video applications. Agi and Gong [18] showed that this algorithm has low security since encrypting of only I-frames offer limited security because of the intercorrelation of frames; some blocks are intracoded in P and B frames. Furthermore, P- and B-frames are highly correlated when they correspond to the same I-frame. They also underlined that it is unwise to encrypt stream headers since they are predictable and can be broken by plaintext-ciphertext pairs. Alatter and Al-ragib [19], apparently unaware of Agi and Gong work [18], stressed the same security leakage. Encryption is performed after compression, thus no impact is observed on the compression efficiency. The resulting bitstream is not MPEG compliant.

### 3.3.3 Methodology proposed by Shi and Bhargava.

In [20], the authors proposed video encryption algorithm (VEA) which uses a secret key to randomly change the signs of all DCT coefficients in an MPEG stream. It is fast as it operates on a small portion of original video. It is more efficient than DES algorithm because it only selectively encrypts a small number of bits of the MPEG compressed video and selected bit is only XORed one time with the corresponding bit of the secret key. VEA does not protect from plaintext attack provided the attacker knowns the original video image (plaintext and ciphertext).

In [21], the authors present a new version of VEA reducing computational complexity; it encrypts the sign bits of differential values of DC coefficients of I-frames and sign bits of differential values of motion vectors of B- and P-frames. This type of improvement makes the video playback more random and more non viewable. When the sign bits of differential values of motion vectors are changed, the directions of motion vectors change as well. In addition, the magnitude of motion vectors change, making the whole video very chaotic. Modified VEA encrypt DC coefficients of I frame, and leave AC coefficients of I frames unchanged. Thus it significantly reduces encryption computations. Because DC coefficients of I frames are differentially encoded, changing a few sign bits of differential values of DC coefficients will affect many DC coefficients during MPEG decoding. MPEG's differential code of DC coefficients and motion vectors increase the difficulty to break MVEA encrypted videos. The first version of VEA [21] is only secure if the secret key is used once. Otherwise, knowing one plaintext and the corresponding ciphertext, the secret key can be computed by XORing the DCT sign bits. Both versions of VEA are vulnerable to chosen plaintext attacks; in [21], it is feasible to create a repetitive/periodic pattern and then compute its inverse DCT. The encryption of the image obtained will allow us to get the key length and even compute the secret key by chosen-plaintext attack.

### 3.3.4 Methodology proposed by Shi, Wang and Bhargava

In [22], a new version of the modified VEA presented in [21] is proposed, called real time video encryption algorithm (RVEA). It encrypts selected sign bits of the DC coefficients and/or sign bits of motion vectors using DES or IDEA. It selects atmost 64 sign bits from each macroblock. RVEA achieves the goal of reducing and bounding its computation time by limiting the maximum number of bits selected. The differential encoding of DC coefficients and motion vectors in MPEG compression increases difficulty of breaking RVEA encrypted videos. If the initial guess of a DC coefficient wrong, it is very difficult to guess the following DC values correctly.

### 3.3.5. Methodology proposed by Wu and Kuo

In [23],[24], based on a set of observations, the authors point out that energy concentration does not mean intelligibility concentration. Indeed, they discussed the technique proposed by Tang [11]. They show that by fixing DC values at a fixed value and recovering AC coefficients (by known or chosen plaintext attacks), a semantically good reconstruction of the image is obtained. Even using a very small fraction of the AC coefficients does not fully destroy the image semantic content. The authors argued that both orthogonal transform-based compression algorithms followed by quantization and compression algorithms that end with an entropy coder stage are bad candidates to selective encryption. They investigate another approach that turns entropy coders into ciphers. They propose two schemes for the most popular entropy coders: multiple Huffman tables (MHTs) for the Huffman coder and multiple state index (MSI) for the QM arithmetic coder.

**MHT**: The authors propose a method using multiple Huffman coding tables. The input datastream is encoded using multiple Huffman tables. The content of these tables and the order that they are used are kept secret as the key for decryption. In the proposed system, instead of training thousands of Huffman coding tables, it only train and obtained four different Huffman tables. Then, thousands of different tables can be derived using a technique called Huffman tree mutation. Gillman and Rivest [25] showed that decoding a Huffman coded bit stream without any knowledge about the Huffman coding tables would be very difficult. However, the basic MHT is vulnerable to known and chosen plaintext attacks as pointed out in [26].

**MSI**: The arithmetic QM coder is based on an initial state index; the idea is to select 4 published initial state indices and to use them in a random but secret order. Unlike Huffman coding with a fixed and pre defined Huffman tree, the QM coder dynamically adjusts the underlying statistical model to a sequence of received binary symbols. It is very difficult to decode the bitstream without the knowledge of the state index used to initialize the MQcoder. A little effect on compression efficiency is observed. This is due to multiple initializations of the QM coder due to initial state index changing.

### 3.3.6 Methodology proposed by Wen, Severa, Zeng, Luttrel, and Jin

A general selective encryption approach for fixed and variable length codes (FLC and VLC) is proposed in [27]. FLC and VLC codewords corresponding to important information carrying fields are selected. Then, each codeword in the VLC and FLC (if the FLC code space is not full) table is assigned a fixed length code index, when we want to encrypt the concatenation of some VLC (or FLC) codewords, only the indices are encrypted (using DES). Then the encrypted concatenated indices are mapped back to a different but existing VLC. The encryption process compromises the compression efficiency. Indeed, some short VLC codewords (which are the most probable/frequent) can be replaced by longer ones. This is antagonistic with the entropy coding idea. The proposed scheme is fully compliant to any compression algorithm that uses VLC or FLC entropy coder.

### 3.3.7. Methodology proposed by Zeng and Lei

In [28], selective encryption in the frequency domain (8 × 8 DCT and wavelet domains) is proposed. The general scheme consists of selective scrambling of coefficients by using different primitives such as selective bit scrambling, block shuffling, and/or rotation. In wavelet transform case selective bit scrambling and block shuffling is done. In selective bit scrambling the first nonzero magnitude bit and all subsequent zero bits if any give a range for the coefficient value. These bits have low entropy and thus highly compressible and all remaining bits called refinement bits are uncorrelated with the neighbouring coefficients. In this scheme, signbits and refinement bits are scrambled. In block shuffling, the basic idea is to shuffle the arrangement of
coefficients within a block in a way to preserve some
spatial correlation; this can achieve sufficient security without compromising compression efficiency. Each subband is split into equal-sized blocks. Within the same subband, block coefficients are shuffled according to a shuffling table generated using a secret key. Since the shuffling is block based, it is expected that most 2D local subband statistics are preserved and compression not greatly impacted.

In DCT transform case, the 8 × 8 DCT coefficients can be considered as individual local frequency components located at some subband. The block shuffling and sign bits change can be applied on these "subbands." I, B, and P frames are processed in different manners. For I-frames, the image is first split into segments of macroblocks, blocks/macroblocks of a segment can be spatially disjoint and chosen at random spatial positions within the frame. Within each segment, DCT coefficients at the same frequency location are shuffled together. Then, sign bits of AC coefficients and DC coefficients are randomly changed. There may be many intracoded blocks in P- and B-frames. At least DCT coefficients of the same intracoded block in P- or B-frames are shuffled. Sign bits of motion vectors are also scrambled. It is vulnerable to chosen and known plaintext attacks since it is based only on permutations. In addition, replacing the DC coefficients with a fixed value still gives an intelligible version of the image. This algorithm can be part of permutation based encryption.

### 3.3.8 Methodology proposed by Bergeron and Lamy-Bergot.

A syntax compliant encryption algorithm is proposed for H.264/AVC [29]. Encryption is inserted within the encoder. Using the proposed method allows to insert the encryption mechanism inside the video encoder, providing a secure transmission which does not alter the transmission process. The bits "selected for encryption" are chosen with respect to the considered video standard according to the following rule: each of their encrypted configurations gives a non-desynchronized and fully standard compliant bitstream. This can in particular be done by encrypting only parts of the bitstream which have no or a negligible impact in evolution of the decoding process, and whose impact is consequently purely a visual one.

About 25% of I-slices and 10–15% of P-slices are encrypted. Since intracoded slices can represent 30–60%, the encryption ratio is expected to be relatively high. The main drawback of this scheme is the lack of cryptographic security. Indeed, the security of the encrypted bitstream does not depend more on the AES cipher. It depends on the size of the compliant codewords. Hence, the diffusion of the AES cipher is reduced to the plaintext space size. In addition, a bias is introduced in the ciphertext. This bias depends on the key size and the plaintext space size.

### 3.3.9 Methodology proposed by Lian, Liu, Ren and Wang.

This scheme is proposed for AVC [30]. During AVC encoding, such sensitive data as intra prediction mode, residue data and motion vector are encrypted partially. Among them, intra prediction mode is encrypted based on exp-golomb entropy coding, the intra macroblocks DCs are encrypted based on context based adaptive variable length coding, and intra macroblocks ACs and the inter macroblocks MVDs are sign encrypted with a stream cipher followed with variable length coding. The encryption scheme is of high key sensitivity, which means that slight difference in the key causes great differences in cipher video and that makes statistical or differential attack difficult. It is difficult to apply known plaintext attack. In this encryption scheme, each slice is encrypted under the control of a 128 bit sub-key. Thus, for each slice, the brute force space is $2^{128}$; for the whole video, the brute force space is $2^{256}$ (the user key is of 256 bit). This brute force space is too large for attackers to break the cryptosystem. According to the encryption scheme proposed here, both the texture information and the motion information are encrypted, which make it difficult to recognize the texture and motion information in the video frames.

## 3.4 Perceptual Encryption

In many applications like pay-per-view video, pay TV and video on demand perceptual encryption is useful. This feature requires that quality of audio and visual data is only partially degraded by encryption i.e. the encrypted multimedia data are still partially perceptible after encryption. Such perceptibility makes it possible for potential users to listen/view low quality versions of the multimedia products before buying them. It is desirable that the aural/visual quality degradation can be continuously controlled by a factor p, which generally denotes a percentage corresponding to the encryption strength [31].

### 3.4.1 Methodology proposed by Pazarci-Dipcin

Pazarci and Dipcin [32] proposed an MPEG-2 perceptual encryption scheme, which encrypts the video in the RGB color space via four secret linear transforms before the video is compressed by the MPEG-2 encoder. The proposed solution is an MPEG-2 transparent video scrambling that allows the unauthorized users to have an arbitrarily degraded view of the current program. The scrambling is performed prior to the MPEG encoding, and the scramble video is MPEG-2 encoded with only a minimal increase in the MPEG transport stream bitrate. The main merit of the Pazarci-Dipcin scheme is that the encryption/decryption and MPEG encoding/decoding processes are separated, which means that encryption part can simply be added to an MPEG system without any modification. However, the following defects make this scheme problematic in real applications. This is because the motion compensation algorithm may fail to work for encrypted videos. The main reason is that the corresponding SBs may be encrypted with different parameters. To reduce this kind of influence, the encryption parameters of all SBs have to be sufficiently close to each other. This However compromises the encryption performance and the security. Unrecoverable quality loss caused by the encryption always exists. Even authorized users who know the secret key cannot recover the video with the original quality [32]. The scheme is not secure against brute force attacks as for a given color component C of any 2X2 SB structure, one can exhaustively guess the alpha values of the four SBs to recover 2X2 SB structure by minimizing the block artifacts occurring between adjacent SBs. The scheme is not sufficiently sensitive to the mismatch of the secret key, since the encryption transforms and the alpha rule given in [32] are both linear functions. The scheme is not secure enough against known/chosen plaintext attacks. This is because value of alpha can be derived approximately from the linear relation between the plain pixel values and the cipher pixel values in the same SB.

### 3.4.2 Methodology proposed by Lian, Wang, Sung and Wang

This scheme is proposed for 3D-SPIHT compressed videos [33-35]. In this scheme confusing different number of wavelet coefficients, encrypting different number of coefficients signs and confusing positions of different data cubes, videos can be degraded to different degrees under the control of quality factor. Its encryption strength can be adjusted according to certain quality factor. It is not secure against known chosen plaintext attack.

### 3.4.3 Methodology proposed by Wang, Yu and Zheng

A different scheme working in DCT domain was proposed by Wang, Yu and Zheng [36]. In this scheme, three new parameters $k_1$, $k_2$, $k_3$ are introduced to determine the values of $a_i$ for three color components, Y, $C_b$, $C_r$. The 16 average values from $a_0$ to $a_{15}$, the two control factors, beta and C and the three parameters $k_1$, $k_2$, $k_3$ altogether serve as the secret scrambling parameters of each SB. Three different ways are suggested for the transmission of secret parameters: a) encrypting them and transmitting them in the payload of transport stream; b) embedding them in the high frequency of DC coefficients c) calculating them from previous I frame. Though the reduction of the compression ratio about motion compensations is avoided, the encryption will change the natural distribution of the DCT coefficients and thus reduce the compression efficiency of the Huffman entropy coder. In addition if the secret parameters are embedded into high quality frequency of DCT coefficients for transmission, the compression performance will be further compromised. The scheme is still not sufficiently sensitive to the mismatch of secret parameters, since the encryption function and the calculation function of $a_i$ are kept linear. It is not sufficiently secure against brute force attacks to the secret parameters because of the limited values of $a_i$, beta, C, $k_1$, $k_2$, $k_3$. Furthermore, due to the non-uniform distribution of the DCT coefficients in each sub-band, an attacker needs not to randomly search all possible values of ai. This scheme is still insecure against known/chosen plaintext attacks if the third way is used for calculating the secret parameters. In this case, $a_i$ of each SB can be easily calculated from the previous I-frame of the plain video. Similarly Beta and C value cab be derived approximately.

### 3.4.4 Methodology proposed by Li, Chen, Cheung, Bharat Bhargava, and Kwok-Tung Lo

This design is a generalized version of VEA for perceptual encryption, by selectively encrypting FLC data elements in the video stream [31]. Apparently, encrypting FLC data elements is the most natural and perhaps the simplest way to maintain all needed features, especially the need for strict size preservation. To maintain format compliance, only last four FLC data elements are considered, which are divided into three categories i) intra DC coefficient ii) sign bits of non intra DC coefficients and AC coefficients iii) sign bits and residuals of motion vectors. Based on this division, three control factors $P_{sr}$, $P_{sd}$ and $P_{mv}$ in the range [0, 1] are used to control the visual quality in three different dimensions like low resolution spatial view, high resolution spatial details and the temporal motions. Known and chosen plaintext attack is ensured by four different measures by implementing block cipher, using a cipher with plaintext/ciphertext feedback, using a key management system and a stream cipher and using a stream cipher with unique ID.

## 4. Comparison of Video Encryption Algorithms

In this section, we present comparison of these algorithms with respect to various parameters as shown in Table 1. Table 1 summarized the related work with each comparison criteria explained in section 2.2. The main symbols used are

a. "H" for High,
b. "V" for variable,
c. "L" for Low,
d. ? for non specified,
e. "√" for satisfied,
f. "X" for not satisfied.

*Table1. Comparison of Video Encryption Algorithms*

| Algorithm Class | Algorithm/Author Name | VD | ER | S | CS | CF | FC |
|---|---|---|---|---|---|---|---|
| Fully Layered Encryption | Naïve Technique[2] | H | 100% | slow | √ | √ | ? |
| Permutation based Encryption | pure permutation[10] | H | 100% | fast | X | √ | ? |
| | Zig-Zag Permutation[11] | ? | 100% | fast | X | √ | ? |
| | Huffman Codeword Permutation[12] | ? | ? | fast | X | X | ? |
| | Compression logic based Random Permutation[13] | V | V | fast | ? | √ | √ |
| | Correlation Preserving Permutation[14] | ? | ? | fast | √ | √ | √ |
| Selective Encryption | Meyer and Gadegast[15][16] | V | V | V | V | √ | X |
| | Spanos and Maples [17][18][19] | ? | H(30 - 60%) | ? | X | √ | X |
| | Shi and Bhargava [20][21] | H | ? | fast | X | √ | √(MPEG) |
| | Shi, Wang and Bhargava [21][22] | H | ? | fast | X | X | √(MPEG) |
| | Wu and Kuo [23][24][25] (MHT) | H | V | ? | X | √ | X |
| | Wu and Kuo [23][24][25](MSI) | H | L | ? | √ | √ | X |
| | Wen, Severa, Zeng, Luttrel, and Jin[27] | H | <15% | ? | ? | X | √(VLC and FLC) |
| | Zeng and Lei[28] (DWT) | H | 20% | ? | X | √ | √( DWT) |
| | Zeng and Lei[28] (DCT) | H | 20% | ? | X | √ | √( JPEG-MPEG) |
| | Bergeron and Lamy-Bergot[29] | H | X | ? | X | √ | √ |
| | Lian, Liu, Ren and Wang [30] | H | ? | fast | √ | ? | √ |
| Perceptual Encryption | Pazarci-Dipcin [32] | V | V | V | X | X | √ |
| | Lian, Wang , Sung and Wang [33-35] | V | V | V | X | √ | √ |
| | Wang, Yu and Zheng[36] | V | V | V | X | ? | √ |
| | Li, Chen, Cheung, Bhargava, Kwok-Tung Lo [31] | V | V | fast | √ | ? | √ |

## 5. Conclusion

Although an important and rich variety of video encryption algorithms have been proposed in literature, most of the algorithms defined in Table 1 are not secure against cryptanalysis attack. Naïve algorithm provides highest level of security but it is very slow in nature and cannot be used in real time. Permutation based algorithms are generally faster but they do not provide sufficient level of security. Selective encryption algorithms reduces computational complexity by selecting only a minimal set of data to encrypt but their security and speed level generally vary based on which and how many parameters they encrypt. Perceptual encryption algorithms are suitable for application like pay per view TV, video on demand where potential users like to see low quality video before buying them. So, these algorithms are not suitable for applications which demand high security. It is difficult for a single algorithm to satisfy all performance parameters. So, selection of encryption algorithm always depends on requirements of application in use. By looking at table we can conclude that it is a challenge for researchers to design an encryption algorithm which maintains tradeoff among all parameters like visual degradation, speed, encryption ratio, compression friendliness, format compliance and cryptographic security.

## References

[1] Shiguo Lian, Dimitris Kanellopoulos, and Giancarlo Ruffo, "Recent Advances in Multimedia Information System Security," International Journal of Computing and Informatics, Vol. 33, No.1, 2009, pp. 3-24.
[2] Shiguo lian, Multimedia Content Encryption: Algorithms and Application, CRC Press, 2008.
[3] B. Furht, D. Socek, and A. M. Eskicioglu, "Fundamentals of Multimedia Encryption Techniques," Multimedia Security Handbook, CRC Press, 2005.
[4] ISO/IEC 13818:1996, Coding of Moving Pictures and Associated Audio (MPEG-2); Part 1: systems, Part2 : video.
[5] Mitchell J. L., Pennebaker W.B., fogg C.E. and LeGall D.J., MPEG Video Compression Standard, Chapman & Hall, 1996.
[6] ITU-T Rec. H.264/ISO/IEC 11496-10. Advanced Video Coding. Final Committee draft, Document JVT-E022, 2002.
[7] Iain E G Richardson. H.264/MPEG Part10, 2002. Available: http://www.vcodex.com.
[8] NIST: Data Encryption Standard, FIPS 46-3, 1999.
[9] NIST: Advanced Encryption Standard, FIPS 197, 2001.
[10] Adam J. Slaggel. Known-Plaintext Attack Against a Permutation Based Video Encryption Algorithm. Available: http://eprint.iacr.org
[11] L.Tang, "For Encrypting and Decrypting MPEG Video Data Efficiently", in Proceedings of the Forth ACM International Multimedia Conference, 1996, pp. 219-230.
[12] C. Shi and B. Bhargava, "Light Weight MPEG video Encryption Algorithm", in Proceeding of the International Conference on Multimedia, 1998,pp. 55-61.
[13] Hao Wang and Chong-wei Xu, "A New Lightweight and Scalable Encryption Algorithm for Streaming Video over Wireless Networks", International Conference on Wireless Network, 2007, pp. 180-185.
[14] D. Socek, H. Kalva, S. S. Magliveras, O. Marques, D. Culibrk, and B. Furht, "A Permutation-based Correlation-Preserving Encryption Method for Digital Videos," in International Conference on Image Analysis and Recognition, 2006,LNCS 4141,pp. 547-558.
[15] J. Meyer and F. Gadegast, "Security Mechanisms for Multimedia Data with the Example MPEG-1 video," Project Description of SECMPEG, Technical University of Berlin, 1995.
[16] L. Qiao and K. Nahrstedt, "A New Algorithm for MPEG Video Encryption," in Proceedings of the 1st International Conference on Imaging Science, Systems and Technology, 1997.
[17] G.A. Spanos and T.B.Maples, "Performance Study of a Selective Encryption Scheme for the Security of Networked Real Time Video," in Proceedings of the International Conference on Computer Communications and Networks,1995, pp. 2-10.
[18] I. Agi and L. Gong, "An Empirical Study of Secure MPEG Video Transmissions," in Proceedings of the Symposium on Network and Distributed System Security, 1996, pp. 137-144.
[19] A.M. Alatter and G.I.Al-regib, "Evaluation of Selection Algorithms for Secure Transmission of MPEG Compressed Bit streams," in Proceedings of IEEE International Symposium on Circuits and Systems, 1999,vol. 4, pp. 340-343.
[20] C. Shi and B. Bhargava, "A Fast MPEG Video Encryption Algorithm," in Proceedings of the 6th ACM International Conference on Multimedia, 1998, pp. 81–88.
[21] C. Shi and B. Bhargava, "An efficient MPEG video encryption algorithm," in Proceedings of the 17th IEEE Symposium on Reliable Distributed Systems 1998, pp. 381–386.
[22] C. Shi, S. Y.Wang, and B. Bhargava, "MPEG Video Encryption in Real-time using Secret Key Cryptography," in Proceedings of the International Conference on Parallel and Distributed Processing Algorithms and Applications, 1999,pp. 191–201.
[23] C.-P. Wu and C.-C. J. Kuo, "Fast Encryption Methods for Audiovisual Data Confidentiality," in Proceedings of SPIE, 2001, Vol. 4209, pp. 284–295.
[24] C.-P. Wu and C.-C. J. Kuo, "Efficient Multimedia Encryption via Entropy Codec Design," in Security and Watermarking of Multimedia Contents ,in Proceedings of SPIE,2001, Vol. 4314, pp. 128–138.
[25] D. W. Gillman and R. L. Rivest, "On breaking a Huffman code," IEEE Transactions on Information Theory, vol. 42, no. 3, 1996,pp. 972–976.
[26] J. Zhou, Z. Liang, Y. Chen, and O. C. Au, "Security Analysis of Multimedia Encryption Schemes based on Multiple Huffman Table," IEEE Signal Processing Letters, vol. 14, No. 3, 2007, pp. 201– 204.
[27] J. Wen, M. Severa, W. Zeng, M. H. Luttrell, and W. Jin, "A Format-Compliant Configurable Encryption Framework for Access Control of Video," IEEE Transactions on

Circuits and Systems for Video Technology, Vol. 12, No. 6, 2002, pp. 545–557.

[28] W. Zeng and S. Lei, "Efficient Frequency Domain Selective Scrambling of Digital Video," IEEE Transactions on Multimedia, Vol. 5, No. 1, 2003, pp. 118–129.

[29] C. Bergeron and C. Lamy-Bergot, "Compliant Selective Encryption for H.264/AVC Video Streams," in Proceedings of the 7th IEEE Workshop on Multimedia Signal Processing ,2005, pp. 1–4.

[30] Shiguo Lian, Zhongxuan Liu, Zhen Ren and Haila Wang, "Secure Advanced Video Coding Based on Selective Encryption Algorithms," IEEE Transaction on Consumer Electronics, Vol. 52, No. 2 ,2006, pp. 621-629.

[31] Shunjun Li, Guanrong Chen, Albert Cheung, Bharat Bhargava, and Kwok-Tung Lo, "On the Design of Perceptual MPEG Video Encryption Algorithm", IEEE Transactions on Circuits and Systems for Video Technology, vol. 17, No. 2, 2007, pp. 214-223.

[32] M. Pazarci and V. Dipcin, "A MPEG2-Transparent Scrambling Technique", IEEE Transactions on Consumer Electronics, Vol. 48, No. 2, 2002, pp. 345-355.

[33] S. Lian, X. Wang, J. Sun, and Z. Wang, "Perceptual Cryptography on Wavelet Transform Encoded Videos, "in Proceedings of IEEE International. Symposium on Intelligent Multimedia, Video and Speech Processing, 2004, pp. 57-60.

[34] S. Lian, J. Sun, and Z. Wang, "Perceptual Cryptography on SPIHT Compressed Images and Videos, "in Proceedings of IEEE International Conference on Multimedia and Expo,2004, Vol. 3, pp. 57-60.

[35] S. Lian, "Perceptual Cryptography on JPEG2000 Compressed Images or Videos, " in Proceedings of International Conference on Computer and Information Technology, IEEE Consumer Society, 2004, pp. 78-83.

[36] C. Wang, H.-B Yu, and M. Zheng, "A DCT based MPEG-2 Transparent Scrambling Algorithm", IEEE Transactions on Consumer Electronics, Vol. 49, No. 4, 2003,pp. 1208-1213.

**Jolly Shah** She completed her undergraduate degree in Information Technology at Dharamsinh Desai University in 2004. She obtained her master degree in Computer Engineering from Bharati Vidyapeeth University in 2008. She is pursuing PhD in computer Science from Jaypee Institute of Information Technology. Her research interest are in cryptography, video processing.

**Dr. Vikas Saxena** He completed his undergraduate degree in Computer Science at Rohilkhand Univ. He obtained his master degree from VJIT, Mumbai. He completed his PhD in Computer Science from Jaypee Institute of Information Technology. He is recently serving as an assistant professor in JIIT. He published various papers in International journals and Conferences. His area of interest is Image Processing, Graphics, and Software Engineering.